\documentclass[11pt]{article}
 
\usepackage{graphicx}
\usepackage{color}
\usepackage{comment}
\usepackage{subfig}
\usepackage[labelfont=bf,labelsep=period,justification=raggedright]{caption}

\usepackage{amsmath,amssymb}
\usepackage{algorithm}
\usepackage{algorithmic}
\usepackage{etoolbox}

\newcommand{\algorithmicdoinparallel}{\textbf{do in parallel}}
\makeatletter
\AtBeginEnvironment{algorithmic}{%
  \newcommand{\FORP}[2][default]{\ALC@it\algorithmicfor\ #2\ %
    \algorithmicdoinparallel\ALC@com{#1}\begin{ALC@for}}%
}
\makeatother
\usepackage{changepage}
\usepackage[utf8x]{inputenc}
\usepackage{textcomp,marvosym}
\usepackage{nameref,hyperref}

\begin{document}




\begin{flushleft}
\centering
{
\Large
\textbf{A GPU-Based Solution to Fast Calculation of Betweenness Centrality on Large Weighted Networks}
}
\\
\vspace{1cm}
Rui Fan$^1$, Ke Xu$^1$ and Jichang Zhao$^{2,\star}$
\\
$^1$State Key Lab of Software Development Environment, Beihang University
\\
$^2$School of Economics and Management, Beihang University
\\
$^\star$Corresponding author: jichang@buaa.edu.cn
\end{flushleft}
\vspace{1cm}


\date{\today}

\begin{abstract}
Recent decades have witnessed the tremendous development of network science, which indeed brings a new and insightful language to model real systems of different domains. Betweenness, a widely employed centrality in network science, is a decent proxy in investigating network loads and rankings. However, the extremely high computational cost greatly prevents its applying on large networks. Though several parallel algorithms have been presented to reduce its calculation cost on unweighted networks, a fast solution for weighted networks, which are in fact more ubiquitous than unweighted ones in reality, is still missing. In this study, we develop an efficient parallel GPU-based approach to boost the calculation of betweenness centrality on very large and weighted networks. Comprehensive and systematic evaluations on both synthetic and real-world networks demonstrate that our solution can arrive the performance of 30x to 150x speedup over the CPU implementation by integrating the work-efficient and warp-centric strategies. Our algorithm is completely open-sourced and free to the community and it is public available through \url{https://dx.doi.org/10.6084/m9.figshare.4542405}. Considering the pervasive deployment and declining price of GPU on personal computers and servers, our solution will indeed offer unprecedented opportunities for exploring the betweenness related problems in network science. 
\end{abstract}



\section{Introduction}
\label{sub:intro}

Recent years, the network science, a multidiscipline research area, is concentrated by researchers from different backgrounds such as computer science, biology and physics. In these studies, betweenness centrality (BC) is always applied as a critical metric to measure nodes or edges' significance~\cite{freeman1977set, barthelemy2004betweenness, PhysRevE.67.017101}. For example, Girvan and Newman developed a community detection algorithm based on edge betweenness centrality~\cite{Girvan11062002}, Leydesdorff applied centrality as an indicator of the interdisciplinarity of scientific journals~\cite{leydesdorff2007betweenness} and Motter and Lai established a model of cascading failures with node load being its betweenness~\cite{PhysRevE.66.065102}. However, the extremely high time and space complexity of calculating betweenness centrality greatly limits its applying on large networks. Before the landmark work of Brandes~\cite{brandes2001faster}, the algorithm for computing betweenness centrality requires $O(n^3)$ time and $O(n^2)$ space. While Brandes reduced the complexity to $O(n+m)$ on space and $O(nm)$ and $O(nm+n^2\log(n))$ on time for unweighted and weighted networks, respectively, where $n$ is the number of vertices and $m$ is the number of edges~\cite{brandes2001faster}. However, this improved algorithm still can not satisfy scientific computation requirements in the present information explosion era as more and more unexpected large networks emerge, such as online social networks, gene networks and collaboration networks. For example, Twitter possesses hundreds of millions active users which construct a huge online social network. However, a weighted network with one million nodes may take about one year to calculate its betweenness centrality using Brandes' algorithm, which is an unbearable cost. Because of this, there is a pressing need to develop faster BC algorithm for explorations of diverse domains.

GPU general computing, which provides excellent parallelization, achieves higher performance compared to traditional CPU sequential algorithms in many issues including network science~\cite{Merrill2015high, Wang2015gunrock, Harish2007accelerating, Cong2005experimental}. CUDA is the most popular GPU-computing framework developed by NVIDIA corporation and some researchers have even parallelized Brandes's algorithm by using it~\cite{Shi2011fast,sariyuce2013betweenness, mcLaughlin2014scalable}. However, previous works concentrated on unweighted networks for simplification, but to our best knowledge, most realistic networks are weighted ones. The most significant difference of BC algorithm on unweighted and weighted networks is the shortest path segment. In weighted networks, Dijkstra algorithm should be used to solve the single source shortest path (SSSP) problem rather than Depth First Search (DFS) algorithm. Many efforts in previous work have been devoted to the GPU version of SSSP problem using the well-known Dijkstra algorithm~\cite{Martin2009cuda, ortega2013new, Delling2011phast, Davidson2014work-efficient}. Although these algorithms have been presented and developed, establishing a parallel version of betweenness centrality algorithm on weighted networks is nontrivial because the original SSSP algorithm have to be modified in many critical points for this task and to our best knowledge, a proper and fast solution is still missing. Aiming at filling this vital gap, we propose a fast solution using CUDA to calculate BC on large weighted networks based on previous GPU BC algorithms and SSSP algorithms in this paper.

To make our algorithm more efficient, we make efforts to optimize it by employing several novel techniques to conquer the influence of irregular network structures. Real-world networks have many characters which could deteriorate the performance of GPU parallelization algorithms. For example, the frontier set of nodes is always small compared to the total number of vertices, especially for networks with great diameters. In the meantime, the majority of nodes do not need to be inspected in each step, hence processing all vertices simultaneously in traditional algorithms is wasteful. McLaughlin and Bader proposed a work-efficient strategy to overcome this problem~\cite{mcLaughlin2014scalable}. Another well-known issue is that the power-law degree distribution in realistic networks brings in serious load-imbalance. Several methods were proposed in previous study to conquer this problem, e.g., Merrill et al. employed edge parallel strategy to avoid load-imbalance~\cite{Merrill2015high} and Hong et al. dealt with this problem by using warp technique~\cite{Hong2011accelerating}. In this paper, we systematically investigate the advantages and drawbacks of these previous methods and implement them in our algorithm to solve the above two problems. Experiments on both real-world and synthetic networks demonstrate that our algorithm outperforms the baseline GPU algorithm significantly. Our main contributions are listed as follows:

\begin{itemize}

\item Based on previous GPU parallel SSSP and betweenness centrality algorithms, we propose an efficient algorithm to calculate betweenness centrality on weighted networks, which achieves 30× to 150× speedup over the best existing CPU algorithm on realistic networks.
\item We compare the traditional node-parallel method to the work-efficient version and the warp-centric method. Experiments on realistic networks and synthetic networks demonstrate that the combination of the two strategies works better than others, which achieves 2.65× speedup over the baseline method on realistic networks.
\item We package our algorithm to a useful tool which can be used to calculate both node and edge betweenness centrality on weighted networks. Researchers could apply this tool to conveniently calculate BC on weighted networks fast, especially on large networks. The source code is publicly available through \url{https://dx.doi.org/10.6084/m9.figshare.4542405}.

\end{itemize}

\section{Background}
First we briefly introduce the well-know Brandes’s algorithm and Dijkstra algorithm based on the preliminary definitions of network and betweenness centrality.

\subsection{Brandes's algorithm}
A graph can be defined as $G(V,E)$, where $V$ is the set of vertices, and $E$ is the set of edges. An edge is a node pair $(u,v,w)$, which means that there is a link connecting nodes $u$ and $v$, and its weight is $w$. If the edge $(u,v)$ exists, it can be traversed from $u$ to $v$ and from $v$ to $u$ because we only focus on undirected graphs in this paper. However, our algorithm can be expanded to directed graph version easily. A path $P=(s,...,t)$ is defined as a sequence of vertices connected by edges, where $s$ is the starting node and $t$ is the end node. The length of $P$ is the sum of the weights of the edges involved in $P$. $d(s,t)$ is the distance between $s$ and $t$, which represents the minimum length of all paths connecting $s$ and $t$. $\sigma_{st}$ denotes the number of shortest paths from $s$ to $t$. According to the definition, we have $d(s,s)=0,$ $\sigma_{ss}=1$, $d(s,t)=d(t,s)$ and $\sigma_{st}=\sigma_{ts}$ for undirected graph. $\sigma_{st}(v)$ denotes the number of shortest paths from $s$ to $t$ where $v$ lies on. Based on these definitions, the betweenness centrality can be defined as

\begin{eqnarray}
\label{eq:bc}
	C_B(v)=\sum_{s\neq v\neq t\in V}\frac{\sigma_{st}(v)}{\sigma_{st}}.
\end{eqnarray}
From the above definitions, the calculation of betweenness centrality can be naturally separated into the following two steps:
\begin{enumerate}
	\item{Compute $d(s,t)$ and $\sigma_{st}$ for all node pairs $(s,t),$}
	\item{Sum all pair-dependencies,}
\end{enumerate}
in which pair-dependency is defined as $\delta_{st}(v)=\frac{\sigma_{st}(v)}{\sigma_{st}}$. The first step consumes $O(mn)$ and $O(mn+n^2\log(n))$ time for unweighted and weighted graph respectively, therefore the bottleneck of this algorithm is the second step, which requires $O(n^3)$ time. Brandse's developed a more efficient BC algorithm which requires $O(mn)$ time for unweighted graph, and $O(mn+n^2\log(n))$ time for weighted graph. The critical point is that the dependency of a node $v$ when the source node is $s$ is $\delta_s(v)=\sum_{u:v\in P_s(u)}\frac{\sigma_{sv}}{\sigma_{su}}(1+\delta_s(u))$. Applying this equation, we can accumulate the dependencies after computing the distance and number of shortest paths from a source vertex $s$ to all other vertices, rather than after computing all pair shortest paths.

We can develop a parallel version based on Brandes's algorithm for unweighted graph because the graph is always traversed as a tree by using DFS algorithm. Given a source node $s$, the root of the tree is $s$ and the tree produced by DFS method in the first step. In the second step, dependencies related to source node $s$ are calculated from the bottom to the root of the tree and the nodes at the same level are isolated and have no influence to each other. As a result, the parallel version can explore nodes at the same level simultaneously in both of the two steps, which will essentially boost the calculation.

\subsection{Dijkstra algorithm}
Dijkstra algorithm~\cite{Dijkstra1959note} and Floyd-Warshall algorithm~\cite{Floyd1962algorithm} are commonly employed to solve shortest path problems. While Dijkstra algorithm is more adaptable to betweenness centrality problem because Brandes's algorithm accumulates dependencies after computing single source shortest paths (SSSP), rather than finding and storing all pair shortest paths. Dijkstra algorithm applies greedy strategy to solve SSSP. In this algorithm, the source node is $s$ and if the shortest path from $s$ and another node $u$ is achieved, $u$ will be settled. According to be settled or not, all nodes in graph $G$ could be separated into two sets, which are settled vertices $S$ and unsettled vertices $U$. An array $D$ is used to store tentative distances from $s$ to all nodes. At first, $D(s)=0$ and $D(u)=\infty$ for all other nodes. And the source node $s$ is settled and considered as the frontier node to be explored. In the second step, for every node $u\in U$ and the adjacent frontier node $f$, if $D[f]+w(f,u)<D[u]$, $D[u]$ will be updated to $D[f]+w(f,u)$. Then the node $v\in U$ that has the smallest distance value will be settled and considered as the new frontier node and then the procedure goes back to the second step. The algorithm finishes when all nodes are settled. From the above description, Dijkstra algorithm has no parallel character as it picks one frontier node in each iteration. But this restriction can be loosed that several frontier vertices can be explored simultaneously which is similar to DFS parallel approach.

\section{GPU-based Algorithm}

\subsection{Parallel betweenness centrality algorithm}
In this section, we introduce the details of our GPU version BC algorithm on weighed graph. Firstly, we apply \emph{Compressed Sparse Row} (CSR) format, which is widely used in graph algorithms, to store the input graph~\cite{Bell2009implementing,Davidson2014work-efficient}. It is space efficient that both of the vertex and edge consume one entry, and it is convenient to perform the traversal task on GPU. Moreover, edges related to the same vertex store consecutively in memory which makes warp-centric technique more efficient. For storing weighted graphs, another array that stores the weights of all edges is accordingly required.

We apply both coarse-grained (that one block processes one root vertex $s$) and fine-grained parallel (that threads within the block compute shortest paths and dependencies that related to $s$) strategies. The pseudo-code in this paper describes the parallel procedure of threads within a block. Algorithm~\ref{alg:init} shows the initialization of required variables. $U$ and $F$ represent unsettled set and frontier set, respectively. $v$ is unsettled if $U[v]=1$ and is frontier node if $F[v]=1$. $d$ represents the tentative distance and $\sigma[v]$ is the number of shortest paths from $s$ to $v$. $\delta[v]$ stores the dependencies of $v$. $S$ and $ends$ record the levels of traversal as CSR format and they are used in the dependency accumulation step. As can be seen in Algorithm~\ref{alg:dependency}, in the dependency accumulation part, we get nodes at the same level from $S$ and $ends$ and accumulate dependencies of these nodes simultaneously. Note that in Algorithm~\ref{alg:dependency} we only assign threads for nodes that need to be inspected rather than assign for all nodes, which enhances the efficiency by avoiding redundant threads.

\begin{algorithm}[H]
\caption{Betweenness Centrality: Variable Initialization}
\label{alg:init}
\begin{algorithmic}[1]
	\FORP{$v \in V$}
		\STATE $U[v] \leftarrow 1$\\
		\STATE $F[v] \leftarrow 0$\\
		\STATE $d[v] \leftarrow \infty$\\
		\STATE $\sigma[v] \leftarrow 0$\\
		\STATE $\delta[v] \leftarrow 0$\\
		\STATE $ends[v] \leftarrow 0$\\
		\STATE $S[v] \leftarrow 0$\\
	\ENDFOR
	\STATE $d[s] \leftarrow 0$\\
	\STATE $\sigma[s] \leftarrow 1$\\
	\STATE $U[s] \leftarrow 0$\\
	\STATE $F[s] \leftarrow 1$\\
	\STATE $S[0] \leftarrow s;S_{len} \leftarrow 1$\\
	\STATE $ends[0] \leftarrow 0;ends[1] \leftarrow 1;ends_{len} \leftarrow 2$\\
	\STATE $\Delta \leftarrow 0$
\end{algorithmic}
\end{algorithm}

\begin{algorithm}[H]
\caption{Betweenness Centrality: Shortest Path Calculation by Dijkstra Algorithm}
\label{alg:shortest}
\begin{algorithmic}[1]
	\WHILE{$\Delta < \infty$}
		\FORP{$v \in V$ $\AND$ $F[v] = 1$}
			\FOR {$w \in neighbors(v)$}
				\IF {$U[w] = 1$ \AND $d[v] + weight_{vw} < d[w]$}
					\STATE {$d[w] \leftarrow d[v] + weight_{vw}$}\\
					\STATE {$\sigma[w] \leftarrow 0$}
				\ENDIF
				\IF {$d[w] = d[v] + weight_{vw}$}
					\STATE {$\sigma[w] \leftarrow \sigma[w] + \sigma[v]$}
				\ENDIF
			\ENDFOR
		\ENDFOR
		\STATE {$\Delta \leftarrow \infty$}
		\FORP {$v \in V$}
			\IF {$U[v] = 1$ $\AND$ $d[v] < \infty$}
				\STATE {$atomicMin(\Delta, d[v] + \Delta_v)$}
			\ENDIF
		\ENDFOR
		\STATE {$cnt \leftarrow 0$}
		\FORP {$v \in V$}
			\STATE {$F[v] \leftarrow 0$}\\
			\IF {$U[v] = 1$ $\AND$ $d[v] < \Delta$}
			\STATE {$U[v] \leftarrow 0$}\\
			\STATE {$F[v] \leftarrow 1$}\\
			\STATE {$t \leftarrow atomicAdd(S_{len}, 1)$}\\
			\STATE {$S[t] \leftarrow v$}\\
			\STATE {$atomicAdd(cnt, 1)$}
			\ENDIF
		\ENDFOR
		\IF {$cnt > 0$}
			\STATE {$ends[ends_{len}] \leftarrow ends[ends_{len} - 1] + cnt$}
			\STATE {$ends_{len} \leftarrow ends_{len} + 1$}
		\ENDIF
	\ENDWHILE
\end{algorithmic}
\end{algorithm}

\begin{algorithm}[H]
	\caption{Betweenness Centrality: Dependency Accumulation}
	\label{alg:dependency}
	\begin{algorithmic}[1]
		\STATE {$depth \leftarrow ends_{len} - 1$}
		\WHILE {$depth > 0$} 
			\STATE {$start \leftarrow ends[depth - 1]$}
			\STATE {$end \leftarrow ends[depth] - 1$}
			\FORP {$0 \leq i \leq end - start$}
				\STATE {$w \leftarrow S[start + i]$}
				\STATE {$dsw \leftarrow 0$}
				\FOR {$v \in neighbors(w)$}
					\IF {$d[v] = d[w] + weight_{wv}$}
						\STATE {$dsw \leftarrow dsw + \sigma[w] / \sigma[v] * (1 + \delta[v])$}
					\ENDIF
				\ENDFOR
				\STATE {$\delta[w] \leftarrow dsw$}
				\IF {$w \not = s$}
					\STATE {$atomicAdd(CB[w], \delta[w])$}
				\ENDIF
			\ENDFOR
			\STATE {$depth \leftarrow depth - 1$}
		\ENDWHILE
	\end{algorithmic}
\end{algorithm}

\subsection{Parallel Dijkstra algorithm}
The parallel version of DFS procedure, which is used in BC algorithm for unweighted network, could be modified naturally from its sequential version because vertices located at the same level of the DFS tree can be inspected simultaneously. While for Dijkstra algorithm, picking one frontier node each time makes its parallelization a difficult task. However, this restriction can be relaxed, which means that several nodes could be settled becoming frontier set and be inspected simultaneously in the next step. In this paper, we apply the method described in~\cite{Crauser1998parallelization,ortega2013new}. In this method, $\Delta_{node\ v}=\min(w(v,u): (v,u)\in E)$ is precomputed. Then we define $\Delta_i$ as

\begin{eqnarray}
\label{eq:delta}
	\Delta_i=\min\{(D(u)+\Delta_{node\ u}):u \in U_i\},
\end{eqnarray}
where $D(u)$ is the tentative distance of node $u$, $U_i$ is the unsettled nodes set in iteration $i$. All nodes that satisfy the following condition

\begin{eqnarray}
\label{eq:condition1}
	D(v) \leq \Delta_i
\end{eqnarray}
are settled and become frontier nodes. When applying Dijkstra algorithm in betweenness centrality calculation, the number of shortest paths should be counted. To achieve this goal, the above condition should be modified to
\begin{equation}
\label{eq:condition2}
	D(v)<\Delta_i.
\end{equation}

\begin{figure}[!h]
\centering
\includegraphics[width=0.8\textwidth]{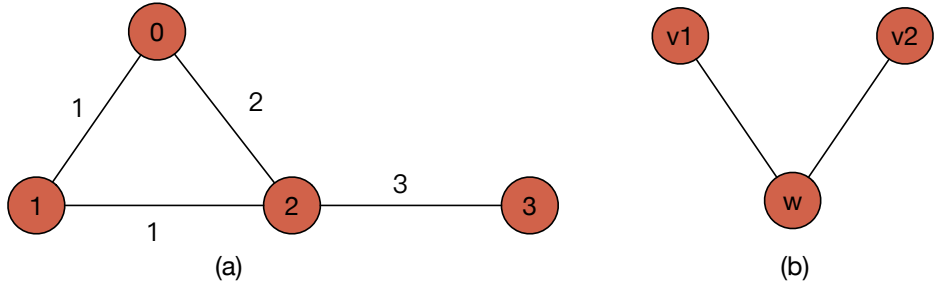}
\caption{(a) An example of choosing frontier nodes, in which using Eq.~\ref{eq:condition1} will make the number of shortest paths of $v_3$ incorrect. (b) An example of race condition. $v_1$ and $v_2$ are frontier nodes in one iteration, and both of which are connected with $w$.}
\label{fig:shortest}
\end{figure}

Fig.~\ref{fig:shortest}(a) demonstrates an example, in which vertex $v_0$ is the source node. If applying Eq.~\ref{eq:condition1}, $v_1$ and $v_2$ will be frontier nodes after inspecting $v_0$ in the first iteration, and the number of shortest paths will be 1 for both $v_1$ and $v_2$. Then $v_1$ and $v_2$ will be inspected simultaneously in next step. If processing $v_2$ first, the number of shortest paths for $v_3$ will be set to 1, while the the correct value of shortest paths' number for $v_3$ should be 2. This mistake comes from the overambitious condition and $v_2$ should not be settled after the first iteration. Although the distance will be correct for all nodes using Eq.~\ref{eq:condition1}, but the number of shortest paths will be wrong. However, Eq.~\ref{eq:condition2} will lead to correct shortest paths number for $v_3$ by only settling $v_1$ after first iteration. This condition could be found at Line 22 in Algorithm~\ref{alg:shortest}.

Algorithm~\ref{alg:shortest} depicts our parallel Dijkstra algorithm in detail. The tentative distance and number of shortest paths are calculated which can be seen from Line 2 to Line 12. In this part, there will be a subtle parallel problem that several nodes in the frontier set may connect to the same node, as can be seen in Fig.~\ref{fig:shortest}(b). In this example, both $v_1$ and $v_2$ are in frontier set and connect to $w$, which results in the classical race condition problem. To avoid this situation, we define a lock for each node. The first thread focus on $w$ will achieve the lock and other threads will not be permitted to change $d[w]$ and $\sigma[w]$. Note that other threads must not wait because in CUDA framework, a group of threads in a warp performs as a SIMD (Single Instruction Multiple Data) unit. After computing $d$ and $\sigma$ for all nodes, we can achieve $\Delta_i$ based on the above analysis, as can be seen from Line 13 to Line 18. In the end, $U$, $F$, $S$ and $ends$ are updated for next iteration.

\subsection{Work-efficient method}
As can be seen on Line 2 in Algorithm~\ref{alg:shortest}, threads will be assigned to all nodes but only nodes that in the frontier set will perform the calculation job, which may be inefficient. McLaughlin et al. figured out an excellent work-efficient technique to solve this problem~\cite{mcLaughlin2014scalable}. Here we develop our work-efficient version by employing this technique. $F$ will be changed to a \emph{queue} that stores all frontier nodes and a variable $F_{len}$ is defined to recode the length of $F$, as can be seen in Algorithm~\ref{alg:we:init}. Then on Line 2 in Algorithm~\ref{alg:we:shortest}, threads can be assigned to $F[0] \sim F[F_{len}-1]$, which may be much smaller than the total number of nodes. At the same time, the method of updating $F$ should also be changed, which can be seen in Algorithm~\ref{alg:we:shortest}.

\begin{algorithm}[H]
\caption{Work-efficient betweenness Centrality: Variable Initialization}
\label{alg:we:init}
\begin{algorithmic}[1]
	\FORP{$v \in V$}
		\STATE // initialize other variables except $F$
	\ENDFOR
	\STATE {$F[0] \leftarrow s$}\\
	\STATE {$F_{len}=1$}\\
	\STATE // initialize other variables
	\end{algorithmic}
\end{algorithm}

\begin{algorithm}[H]
\caption{Work-efficient betweenness Centrality: Shortest Path Calculation by Dijkstra Algorithm}
\label{alg:we:shortest}
\begin{algorithmic}[1]
	\WHILE{$\Delta < \infty$}
		\FORP{$0 \leq i < F_{len}$}
			\STATE $v \leftarrow F[i]$\\
			\STATE // inspect $v$
		\ENDFOR
		\STATE // calculate $\Delta$\\
		\STATE {$F_{len} \leftarrow 0$}\\
		\FORP {$v \in V$}
			\IF {$U[v] = 1$ $\AND$ $d[v] < \Delta$}
			\STATE {$U[v] \leftarrow 0$}\\
			\STATE {$t \leftarrow atomicAdd(F_{len},1)$}\\
			\STATE {$F[t] \leftarrow v$}\\
			\ENDIF
		\ENDFOR
		\IF {$F_{len} > 0$}
			\STATE {$ends[ends_{len}] \leftarrow ends[ends_{len}-1]+F_{len}$}\\
			\STATE {$ends_{len} \leftarrow ends_{len}+1$}\\
			\FOR {$0 \leq i < F_{len}$}
				\STATE {$S[S_{len}+i] \leftarrow F[i]$}
			\ENDFOR
			\STATE {$S_{len} \leftarrow S_{len}+F_{len}$}
		\ENDIF
	\ENDWHILE
\end{algorithmic}
\end{algorithm}

\subsection{Warp-centric method}
Real world networks always have scale-free character, which means their degree distributions follow power law. When implementing parallel graph algorithms through node parallel strategy, this feature brings in serious load-imbalance problem. Most nodes have low degrees while some nodes have extremely high degrees. Threads that assigned to high degree nodes will run slowly and other threads have to wait. Edge parallel strategy can solve this problem~\cite{jia2011edge} but bring in other under-utilizations at the same time. In this paper, we apply the novel warp-centric method~\cite{Hong2011accelerating}, which allocates a warp to one node rather than a thread. Then threads within a warp focus on part of edges connected the specific node. As a result, each thread does less job for high degree nodes and the waiting time will be sharply decreased. Moreover, memory access patterns can be more coalesced than the conventional thread-level task allocation and because of this, the efficiency of memory access can be essentially improved.

\begin{figure}
\centering
\includegraphics[width=0.9\textwidth]{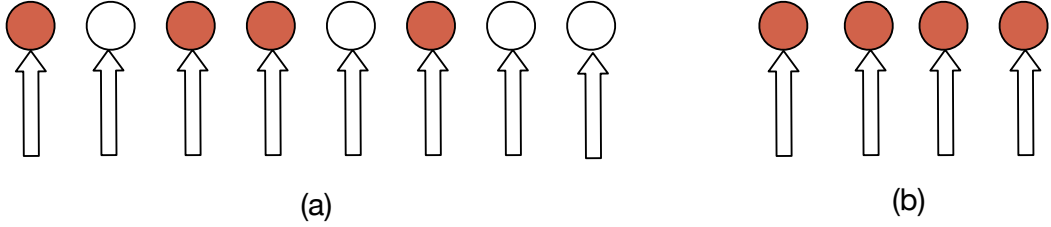}
\caption{An example of threads allocation in node parallel method (a) and work-efficient method (b). Red nodes are frontier nodes that should be processed and an arrow represents a warp that be assigned to the corresponding node. Warp-centric method will waste more threads on nodes that do not need to be inspected. But combining warp-centric and work-efficient method can solve this problem, as shown in (b).}
\label{fig:warp}
\end{figure}

Nevertheless, the warp-centric method also has some drawbacks. Firstly, node degree may be smaller than the warp size, which is always 32 in modern GPU. To solve this problem, virtual warps are proposed in~\cite{Hong2011accelerating}. Secondly, the number of required threads will be raised as each node needs \emph{WARP\_SIZE} threads rather than one thread in this situation. But the number of threads in one block is fixed, hence each thread will be assigned to more nodes iteratively, which may results in low performance. We find that work-efficient method can relieve this problem because it requires less threads compared to the conventional node-parallel method, as can be seen in Fig.~\ref{fig:warp}. In this paper, we apply the warp-centric method for both node-parallel and work-efficient method. As a result, we get four algorithms (see Fig.~\ref{fig:real}) that using different threads allocation strategies and we compare them on both real-world and synthetic networks.

\section{Experiments}

\subsection{Networks and settings}
We collect six real-world networks from the Internet, which have broad types including collaboration network, epinions trust network, email communication network, wiki vote network and two biological networks. They are publicly available in the Internet and have been analyzed extensively by previous literatures~\cite{Richardson2003trust, leskovec2009community, Leskovec2010signed, bansal2007infer, newman2001structure}. The details of these networks are listed in Table~\ref{tab:real}. To further understand the effect of network structures to algorithms' performance, we generate two types of networks, which are Erdős–Rényi (ER) random graphs~\cite{erdos1959on} and Kronecker graphs~\cite{leskovec2010kronecker}. The degree distribution of ER random graph is Poisson, indicating its nodes' degrees are relatively balanced. While Kronecker graph possesses scale-free and small-world characters, which make it more similar to the realistic network. The two biological networks and the cond-mat-2005 collaboration network are weighted networks and for other networks, we uniformly assign random edge weights ranging from 1 to 10. We run the four GPU methods on Geforce GTX 980 (only entry-level for scientific computing) using CUDA 7.5 Toolkit. We also develop the sequential algorithm using C++ and optimize it by applying binary heap in Dijkstra algorithm due to Fibonacci heap's inefficiency in practical use, making our CPU version BC algorithm performs better than most of the existing implementations. And we run the CPU version algorithm on competent Intel Xeon E5620 with 2.40GHz.

\begin{table}[!ht]\footnotesize
\begin{adjustwidth}{-0.6in}{0in}
\centering
\caption{{\bf Details of networks from public dataset}}
\begin{tabular}{|l|l|l|l|l|l|}
\hline

{\bf Network} & {\bf Vertices} & {\bf Edges} & {\bf Max degree} & {\bf Average degree} & {\bf Description}\\ \hline
bio-human-gene1~\cite{nr-aaai15, bansal2007infer} & 22283 & 12345963 & 7940 & 1108.11 & Human gene regulatory network\\ \hline
bio-human-gene2~\cite{nr-aaai15, bansal2007infer} & 14340 & 9041364 & 7230 & 1261.00 & Human gene regulatory network\\ \hline
cond-mat-2005~\cite{newman2001structure} & 39577 & 175693 & 278 & 8.88 & Collaboration network\\ \hline
email enron~\cite{snapnets, leskovec2009community} & 36692 & 183831 & 1383 & 10.02 & Email communication network from Enron\\ \hline
soc epinions1~\cite{snapnets, Richardson2003trust} & 75879 & 405740 & 3044 & 10.69 & Who-trusts-whom network of Epinions.com\\ \hline
wiki vote~\cite{snapnets, Leskovec2010signed} & 7115 & 100762 & 1065 & 28.32 & Wikipedia who-votes-on-whom network\\ \hline
\end{tabular}
\label{tab:real}
\end{adjustwidth}
\end{table}

\subsection{Results}
\begin{figure}
\centering
\includegraphics[width=0.9\textwidth]{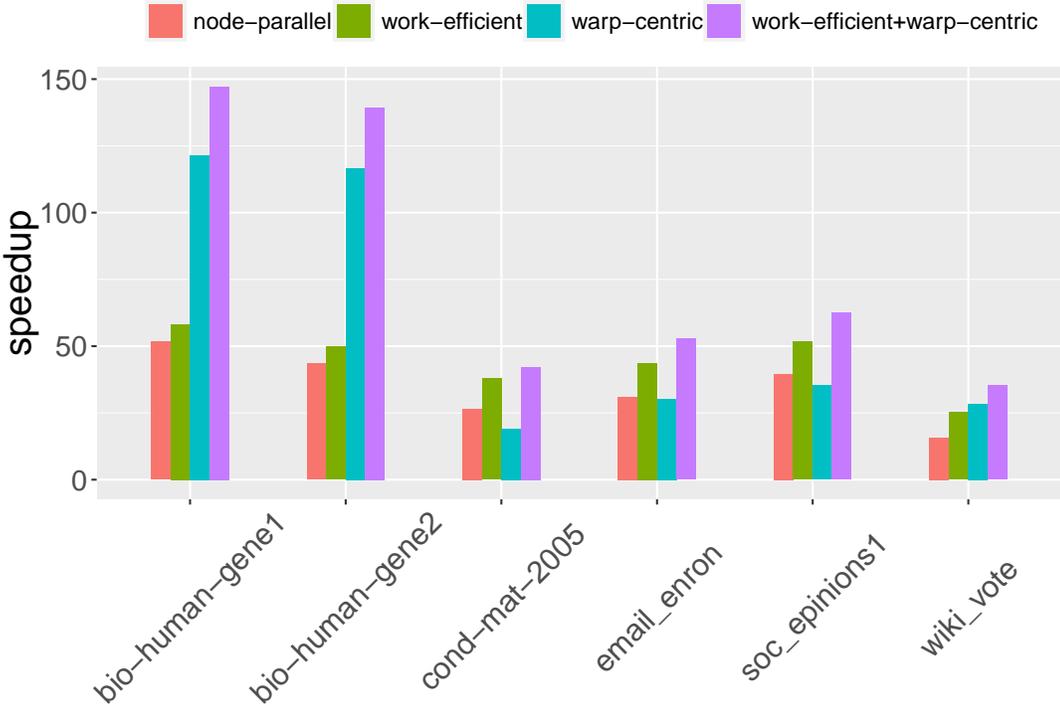}
\caption{Speedups over sequential algorithm of the four GPU implementations on realistic networks. We define speedup as the quotient of the CPU and GPU algorithm running time. The \emph{WARP\_SIZE} is fixed to 32 in the two warp-centric methods.}
\label{fig:real}
\end{figure}

From Fig.~\ref{fig:real}, we can see that all the four GPU programs achieve much better performance than the CPU version on all the six real-world networks. The algorithm that applies work-efficient coupled with warp-centric technique is the best one for achieving 30× to 150× speedup and its performance could be essentially improved on more sophisticated GPU devices. Work-efficient method is more efficient than node-parallel in all networks, while warp-centric method is better on large degree networks, such as the two biological networks. For networks with low average degrees, applying warp-centric method alone is always inefficient because nodes' degrees are always smaller than \emph{WARP\_SIZE}. Using smaller virtual \emph{WARP\_SIZE} could be better on these networks and we will demonstrate this hypothesis later. However, combining warp-centric method and work-efficient method is always better than using work-efficient method alone because it needs less threads in each step, which accordingly relieves the influence of the second drawback of warp-centric method. The combination algorithm achieves 2.65× speedup over the baseline node-parallel strategy on average.

\begin{figure}
\centering
\includegraphics[width=0.9\textwidth]{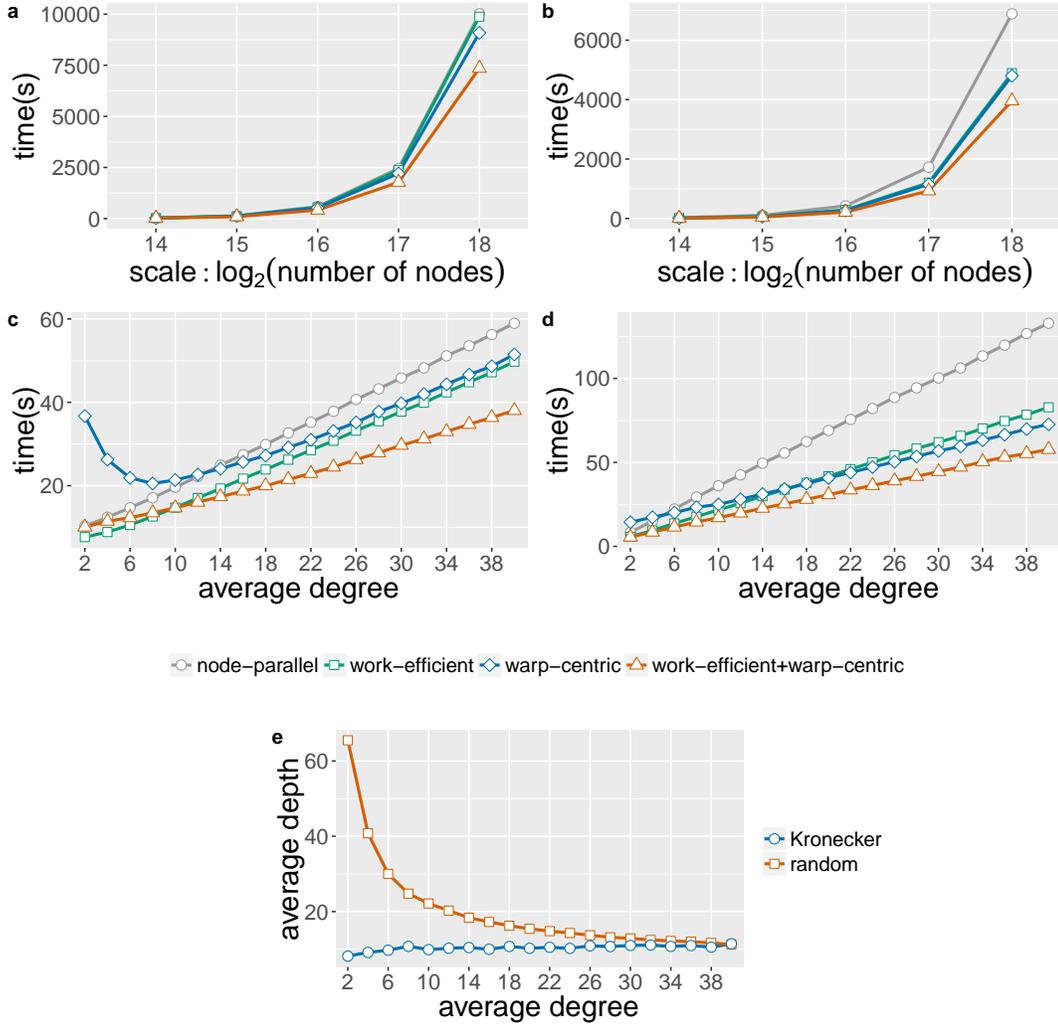}
\caption{Performance of the four implementations on ER random and Kronecker graphs. The \emph{WARP\_SIZE} is fixed to 32 in the two warp-centric methods. (a) and (b) tune the number of nodes from $2^{14}$ to $2^{18}$ for ER random and Kronecker graphs, respectively. And the average degrees are fixed to 32 for both of the two types of networks. (c) and (d) separately tune the average degrees for random and Kronecker networks, in which the random networks have 20000 vertices and the Kronecker networks have $2^{15}$ nodes. (e) illustrates the average depths of search trees for random graphs used in (c) and Kronecker graphs used in (d). Networks with larger depths have smaller average frontier sets, indicating the poor performance with parallelism.}
\label{fig:scale}
\end{figure}

To deeply mining the relationship of the network structure and the performance of the four GPU implementations, we further run them on two types of synthetic graphs, as can be seen in Fig.~\ref{fig:scale}. From Fig.~\ref{fig:scale}(a), (b), (c) and (d), we find that work-efficient algorithm works better than node-parallel algorithm in all networks since it always reduces the required number of threads. As can be seen in Fig.~\ref{fig:scale}(a) and (b), warp-centric method works well on networks with large degrees, which is consistent with the conclusion in realistic networks. Note that for Kronecker graphs, warp-centric method works better than that for random graphs as Kronecker graphs have serious load-imbalance problem and warp-centric technique appropriately solves it. While for ER random graphs in~\ref{fig:scale}(a), the advantage of warp-centric method is only the efficient memory access. For low degree graphs, warp-centric method works even worse than node-parallel strategy because the degrees are always smaller than \emph{WARP\_SIZE}, as can be seen in Fig.~\ref{fig:scale}(c) and (d). For random graphs, the performance of warp-centric method is extremely poor when the average degree is smaller than 8 and Fig.~\ref{fig:scale}(e) explains the reason. The small average degree brings in large average depth, which means that the average size of the frontier sets is small. In this case, warp-centric method assigns more useless threads to nodes that do not need inspections. On the contrary, as the degree grows, it is closer to \emph{WARP\_SIZE} and the depths drop down sharply, which make the warp-centric method performs much better. While low-degree Kronecker graphs have power-law degree distributions and small average depths, which make warp-centric method works not as bad as on random graphs. However, the combination of the two methods always runs faster than applying work-efficient method alone because it avoids the second drawback of warp-centric method, which is discussed in the previous section. In conclusion, work-efficient method always achieves better performance while the performance of warp-centric method relies on networks' structures but the joint version always achieves the best performance.

\begin{figure}
\centering
\includegraphics[width=0.9\textwidth]{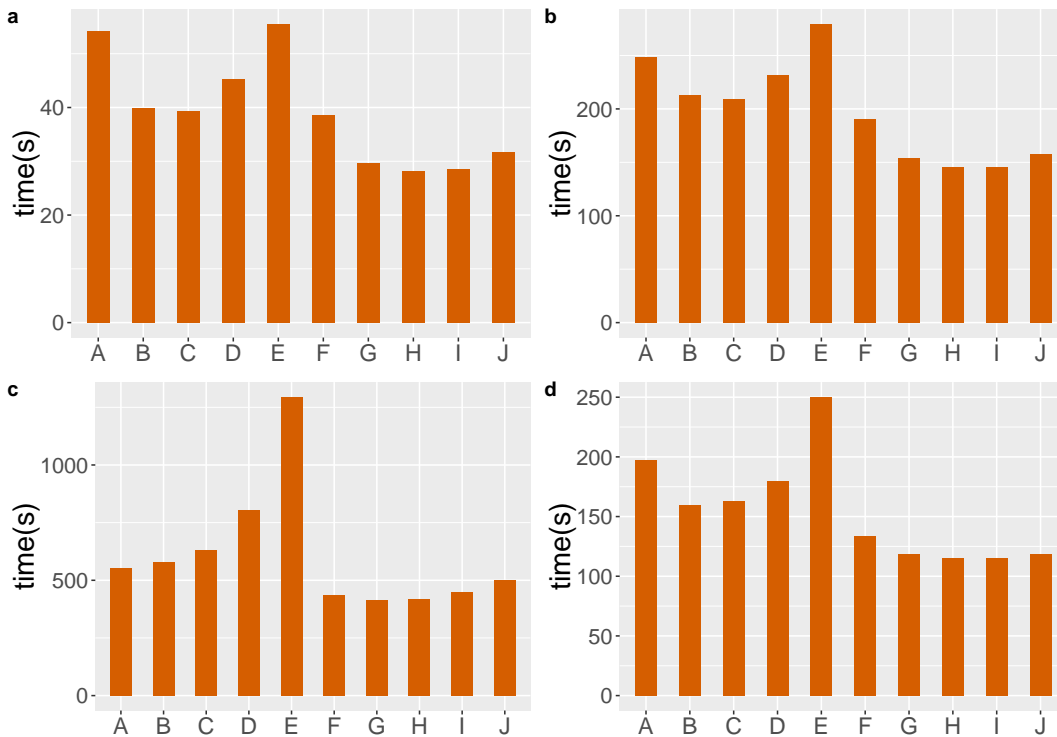}
\caption{Applying other \emph{WARP\_SIZE} on several low-degree networks. A-J represent node-parallel, warp4, warp8, warp16, warp32, work-efficient, work-efficient+warp4, work-efficient+warp8, work-efficient+warp16, work-efficient+warp32, respectively. (a) and (b) are email enron network and soc\_epinions1 network, respectively, on which smaller warp size achieves better performance than both node-parallel method and the algorithm with large \emph{WARP\_SIZE}. (c) is a random graph with $2^{17}$ nodes whose average degree is four. Warp-centric method can not accelerate the speed when combining node-parallel strategy. But when combining small \emph{WARP\_SIZE} with work-efficient method, the performance will be slightly better than applying work-efficient method alone. (d) is Kronecker graph with $2^{17}$ nodes and the average degree is four, on which smaller \emph{WARP\_SIZE} achieves better performance.}
\label{fig:warp_adjust}
\end{figure}

From the above analysis, applying smaller \emph{WARP\_SIZE} may accelerate the two implementations which using warp-centric method when the netowrks' average degree is small. And this hypothesis can be verified in Fig.~\ref{fig:warp_adjust}. We apply smaller \emph{WARP\_SIZE} on email enron network, soc\_epinions1 network and other two synthetic graphs whose average degrees are both four. From Fig.~\ref{fig:warp_adjust}(a) and (b), we find that implementations with smaller \emph{WARP\_SIZE} do perform better than both of the baseline node-parallel algorithms and the large \emph{WARP\_SIZE} algorithm on both of the low-degree realistic networks. And when coupled with work-efficient method, algorithms with smaller \emph{WARP\_SIZE} also perform better than both of the work-efficient strategy alone and the combination of work-efficient and large \emph{WARP\_SIZE}. The reason is that small \emph{WARP\_SIZE} reduces the required number of threads and then eliminates the waste of assigning more threads to a node than its degree. The implementations which have small \emph{WARP\_SIZE} and coupled with work-efficient method achieve the best performance because they avoid both drawbacks of warp-centric method but utilize the advantages of this technique. The results on low-degree Kronecker graph is similar as on realistic networks since Kronecker graph is similar with real-world network. For ER random graphs, the algorithm with small \emph{WARP\_SIZE} does not achieve better performance compared to node-parallel version because the large average depth, which is analyzed in previous section. However, when coupled with work-efficient method, the implementations with small \emph{WARP\_SIZE} perform slightly better than the work-efficient algorithm, which further proves the excellence and stability of the joint algorithm. In summary, the joint algorithm are most efficient and insensitive to network structure. And if we choose an appropriate \emph{WARP\_SIZE}, its performance could be even better.

\section{Conclusion}
Existing GPU version of betweenness centrality algorithms only concentrate on unweighted networks for simplification. Our work that computing betweenness centrality on large weighted networks bridges this gap and achieves prominent efficiency enhancement compared to the CPU implementation. Moreover, we apply two excellent techniques which are work-efficient and warp-centric methods in our algorithm. Work-efficient method allocates threads more efficiently and warp-centric method solves the load imbalance problem and simultaneously optimizes the memory access. We compare these implementations with CPU algorithm and the basic GPU algorithm in realistic networks. The results show that GPU parallel algorithms perform much better than the sequential algorithm and the algorithm which integrates the two techniques is the best, achieving 30× to 150× speedup over the CPU version. Results on generated random graphs and Kronecher graphs further justify the outperformance of our solution.

For future work, we will consider implementing GPU algorithm to process dynamic networks. When networks changes a little (like few new nodes come in or several links vanish), calculating betweenness centrality for all nodes is unnecessary because betweenness centrality of most nodes and edges will not be changed. Some previous works have explored the sequential algorithm on this issue~\cite{Lee2016efficient, singh2015faster, nasre2014betweenness}. We plan to develop GPU version of these algorithms to achieve better performance.

\section*{Acknowledgments}
This work was supported by NSFC (Grant No. 71501005) and the fund of the State Key Lab of Software Development Environment (Grant Nos. SKLSDE-2015ZX-05 and SKLSDE-2015ZX-28). R. F. also thanks the Innovation Foundation of BUAA for PhD Graduates.

\end{document}